\documentstyle[12pt,aasms4]{article}
\begin{document}

\title{Recent Models of Gamma Ray Bursts and Needs for Future Observations}
 
\author{
Igor G. Mitrofanov$^1$
\\[12pt]  
%
$^1$  Space Research Institute, Moscow 117810, Russia,
{\it E-mail: imitrofa@iki.rssi.ru} 
}

\begin{abstract}
The paper presents the nowadays definition of the
phenomenon of cosmic gamma ray bursts, refers to the main
alternative models of their origin and proposes
three promising domains of new observations in the 
incoming decade.
\end{abstract} 


\section{What is the definition of the phenomenon of gamma-ray bursts?}

\vspace{2mm}

Today, 30 years after the detection of the first gamma-ray burst
in July 2, 1967 (Klebesadel {\it et al.} 1973), the definition
of the phenomenon has to be adjusted to the present
observational data. A lot of new features have been recently
discovered, and the phenomenon of {\it classical gamma-ray
burst} (GRB) is becoming to be a rather complex one, which
includes a number of different observational features. To
explain the origin of bursts one has to build the
self-consistent theoretical concept(s), which ought to address
to all these features been taken together.  

Below I present my insight on the basic observational features
of GRBs and on the main theoretical concepts and suggest the
directions of further developments in the field.  

\subsection{Time-related parameters}

\vspace{2mm}

There are four time-related parameters associated with emission
of GRBs. The interval between bursts beginning and end is a {\it
burst duration} $T_{d}$. One can not measure the $T_{d}$ without
many biases. Some of them are associated with the difference of
triggering conditions for bursts with different time histories,
another are related with statistical fluctuations of counts. The
best known version of $T_{d}$ parameter are $T_{50}$ (or
$T_{90}$), which equal to time which takes to accumulate from
25\% (or 5\%) to 75\% (or 95\%) of the total bursts fluence
(Meegan {\it et al.} 1996a). The longest bursts are known to
have $T_{d}$ about several hundreds of seconds, the shortest
events have $T_{d}$ about 30 milliseconds (Fishman and Meegan
1995). Therefore, duration of bursts are distributed over more
than 4 orders of magnitude. 

The bimodal distribution of GRBs was found for duration
parameters (Dezalay {\it et al.} 1992, Kouveliotou {\it et al.}
1993) which probably points out on the existance of different
modes of GRBs with different average values of $T_{50/90}$.
However, parameters $T_{50/90}$ have different physical senses
for bursts with different time profiles: while for single pulse
burst they represent the pulse width, for multi-pulse event
these parameters represent interpulse separations rather than
pulse widths. 

The physical interpretation of {\it duration bimodality} of GRBs
is the important problem of bursts studies.  Another options of
duration parameter have been proposed to study this problem,
which measure either high flux intervals, such as {\it
equivalent pulse width} $T_{epw}$, or low flux interpulses, such
as {\it valley duration} $T_{vd}$, (Mitrofanov {\it et al.}
1997a). It was found that the bimodality is well seen for the
histogram of equivalent pulse width. The two modes have
different time histories, and one might suspect that they
represent two distinct classes of bursts. 

Parameter of {\it flux variations} $T_{fv}$ characterizes the
shortest time scale of bursts emission. For some {\it fast}
bursts it is as short as parts of milliseconds (Bhat {\it et
al.} 1992), for another {\em slow} events, like the {\it FRED}
burst, this time is as long as several seconds (Fishman and
Meegan 1995). The shortest physical value of $T_{fv}$ is not
resolved yet. The observed limit of $T_{fv}$ is thought to be
determined either by 64 ms time resolution of BATSE or by the
poor statistics of counts at short time bins. It seems that all
large number of short bursts are missed at the time scale
shorter then 64 ms. 

Quite recently some evidence for the third characteristic time
of GRBs was found, which is the time of bursts {\it afterglow}
$T_{ag}$. This time is between several hours and tens of days
(see below).  

The longest time scale could be defined, as the possible time of
bursts {\it recurrency} $T_{rec}$. It is not clear yet, are
emitters of bursts really recurrent or not. The most exciting
fact was the BATSE detection of four spatially coincident GRBs
in October 27-29, 1996 (Meegan {\it et al.} 1996b). The rich
statistics of $\sim 2000$ BATSE bursts allows to find the
observational limit for $T_{rec}$: the number of possibly
repeated bursts cannot be larger than about 7\% (Hakkila {\it et
al.} 1996). Therefore, for the total time of 5 years of BATSE
observations, the lowest observational limit for average
$T_{rec}$ is about 40 years.  

\subsection{Photons energy spectra}

\vspace{2mm}

Classical GRBs have variable photons spectra with peak energy
$E_{p}$ of the $\nu F_{\nu}$ curve about several hundreds keV
(Fishman and Meegan 1995). Energy spectrum of a burst varies as
fast as the total flux, and the time scale $T_{fv}$ determins
the variability of energy spectra also. BATSE bursts have smooth
energy spectra with no evidence for line-like features (Briggs
1996a), while lines at low energies $\le 100$~keV are still seen
by some another instruments (Aptekar {\it et al.} 1996). The
problem of {\it no-lines} has to be resolved for further
progress in the bursts studies.  

A group of BATSE bursts was separated, as {\it high emission}
(HE) bursts. They have significant flux at high energies above
$\sim300~$keV.  A complementary group was named, as {\it no high
emission} events (NHE), because they do not have a significant
emission at this range (Pendleton {\it et al.} 1997).
Correspondingly, some BATSE bursts have energy spectra with
exponential cut-off above $\sim 300$~keV, while some another
events have the power low spectra at high energies (Band 1996).
Using the data of instruments COMPTEL and EGRET, the
significant high-energy emission component was found for 5
bright BATSE bursts with energies as high as several GeV (Dingus
1995, Hurley {\it et al.} 1994).  

On the other hand, the excess of soft X-rays at 5-10 keV was
found for $\sim 10\%$ of bright bursts (Preece {\it et al.}
1996). It is well seen above the level of spectral interpolation
from the medium energy range, and it could be interpreted as a
separate emission component.  Further studies has to be done to
find any correlation between emission at different spectral
ranges.  

\subsection{Afterglow emission}

\vspace{2mm}

The first evidence for afterglow emission was found at high
energy gamma rays, when GeV photons were detected during
$\sim5000$~s after the end of GRB 940217 time profile (Hurley
{\it et al.} 1994). Recently the afterglow in X-rays was found
for several bursts due to rapid follow-up observations with high
sensitivity (e.g. Costa {\it et al.} 1997, Piro {\it et al.}
1997a,b, Marshall {\it et al.} 1997). The high accuracy of
locations of X-ray transients allowed to find the possible
optical and radio counterparts of GRBs. They had decreasing
luminosity, and could be associated with afterglow of bursts in
visual light and in radio waves (Groot {\it et al.} 1997, Bond
{\it et al.} 1997, Frail {\it et al.} 1997a). 

The fading time of afterglow $T_{ag}$ is about several hours in
X-rays and about days or tens of days in optics and radio. It is
not clear yet, does the afterglow emission accompanies some
distinct group of bursts, or is it a common signature for all
bursts. However, it seems that some bursts (e.g. GRB 970111) have
not afterglow (Frail {\it et al.} 1997b).

The most exciting recent result is associated with the optical
counterparts of GRB 970228 and GRB 970508. In the second case
the red-shifted absorption lines have been resolved in the
optical spectra of a fading object with $Z_{ab}=0.835$ (Metzger
{\it et al.} 1997). If this object is indeed the burst aftergow,
one gains the conclusive identification of the burst with a
cosmological source with $Z_{em}>0.835$. On the other hand, five
GRBs with the smallest known error boxes were recently examined
with HST. No optical objects were found inside them, which would
be similar to that proposed as a counterpart of GRB 970228
(Scharfer {\it et al.} 1997). So, either a burst GRB 970228 has
the unusual afterglow counterpart, or this identification needs
some further approvals.  

\subsection{Bursts on the sky map and on the brightness scale}

\vspace{2mm}

All detected burst are known to lie on the sky with the perfect
isotropy (Fishman and Meegan 1995). No groups of bursts could be
selected among them which would manifest any significant
deviations from it (Briggs {\it et al.} 1996b). However, another
test of a bright sample of bursts led recently to the conclusion
that they are concentrated toward the Galactic Plane and the
Center (Link and Epstein 1997). Further studies are necessary in
this direction.  

The peak fluxes $F_{max}$ of classical bursts varies in more
than 4 orders of magnitude from $\ge 10^{-3}$~erg cm$^{-2}$ down
to $\le 10^{-7}$~erg cm$^{-2}$. The brightness distribution
$log~N/log~F_{max}$ of the largest sample of 3B catalog (Meegan
{\it et al.} 1996a) significantly deviates from the -3/2 slop,
which corresponds to homogeneous distribution of sources in the
three-dimensional Euclidean space. This deficit of dim bursts is
also seen as decrease of the average parameter
$<V/V_{max}>=0.33\pm0.01$ in respect with the value 0.5, which
is expected for the homogeniouse case (Meegan {\it et al.}
1996a).  

\subsection{Average signatures of bursts emission}

\vspace{2mm}

The large number of $\sim 2000$ BATSE bursts allows to find the
generic signatures of bursts, which represent the basic
properties of their emitters. The first found signature was the
{\it hardness-brightness correlation} (Mitrofanov {\it et al.}
1992a,b, 1996; Paciesas {\it et al.} 1992): brighter bursts were
found to be much harder than dimmer events. This effect was also
well seen, as a strong correlation between bursts peak fluxes
$F_{max}$ and spectra peak energies $E_{p}$ (Mallozzi {\it et
al.} 1995).  

The recent statistical studies of energy spectra of BATSE bursts
have shown that groups of bright and dim bursts have similar
evolution trends on the plane of power index $\alpha$ and peak
energy $E_{p}$, but the trend for the dim group is shifted to
larger power indexes $\alpha$ and smaller $E_{p}$ values
relative to one for the bright group. Therefore, the difference
of average spectral parameters between bright and dim samples is
found to be more complicate than the simple correlation between
$E_{p}$ and brightness (Mitrofanov {\it et al} 1997b).  

The effect of {\it duration-brightness anti-correlation} was
discussed by several authors with quite contradicting
conclusions (Norris {\it et al.} 1994, Mitrofanov {\it et al.}
1996). Bursts are known to have very different time histories,
and one could not test the effect by the direct comparison
between particular events. Special average signatures were
implemented to make the comparison, such as the {\it average
emissivity curve} (Mitrofanov {\it et al.} 1996), or {\it
average autocorrelation function} (in't Zand and Fenimore 1996).
The 3$\sigma$ upper limit of stretching factor of dimmer bursts
in respect with the bright sample was recently estimated as $\le
1.5$ (Mitrofanov {\it et al.} 1997c). 

\section{Origin of GRBs}

\vspace{2mm}

The time of bursts variability is so small, the energy of
bursts' photons are so high and estimated emission energy is so
large, that practically all theoretical models identify bursts
with cataclysms on compact relativistic objects of stellar
masses. The difference between models is in the nature of these
cataclysms.  

\subsection{The Cosmological paradigm}

\vspace{2mm}

Studies of binary radio pulsars led to the conclusion that close
relativistic binaries of compact stars at the late stage of
evolution lose the bound energy and finally fall into the
merging stage, when two compact objects coalesce into a black
hole with emission of $E_{mer}\sim 10^{53}$~erg of energy (e.g.
Paczynski 1986, Piran 1992). For a spiral galaxy like Milky Way
the rate of merging is about $\sim 10^{-4} - 10^{-6}$~yr$^{-1}$
(Narayan {\it et al.} 1991, Jorgensen {\it et al.} 1995). This
rate and the total energy release agree with the needs of the
{\it cosmological model} of GRBs, which put mergers at
cosmological distances with the red-shifts for the most distant
objects $Z_{out}=2-6$.  

A burst is thought to be emitted by the relativistic {\it
fireball}, which expands into the interstellar medium from a
place of energy release. The Lorentz factor of fireball
expansion is about $\Gamma \sim 100-1000$. The forward shock
wave propagates outward and interacts with the interstellar
medium. It is accompanied by reversed wave coming inward
(Meszaros and Rees 1993). The prompt emission of gamma ray
bursts is thought to be emitted either by the synchrotron
radiation or by the inverse Compton scattering at the {\it
external shock} wave regions. Lorentz transformation makes this
emission as hard as observed gamma rays $\sim E_{p}$ and
squeezes it into a time imterval as short as bursts time
duration $\sim T_{d}$. At the late stage of expansion the blast
wave is substantially decelerated by accumulated interstellar
matter, which gives rise to continues afterglow emission. It
shifts from X-rays down to optics and radio, and takes a time of
about $T_{ag}$ for the full disappearing.  

The simple model of external shock wave has the well known
difficulty in explanation of the complexity of burstsŽ time
profiles, which is manifested by fast and sharp variations $\sim
T_{fv}$ between pulse and interpulse intervals. The alternative
model has been implemented to solve the problem, which attribute
pulse structure with emission from {\it internal shocks} waves
inside a fireball (Rees and Meszaros 1994). They are thought to
be ignited by explosions of energy source, which sporadically
take place during a time of burst duration $\sim T_{d}$.  

According to the modern cosmological models, the {\it internal
shock} wave is responsible for the prompt bursts emission, while
the {\it external shock} waves are more likely associated with
the afterglow emission (Katz and Piran 1997).  

\subsection{The Galactic Paradigm}

\vspace{2mm}

The galactic paradigm associate GRBs with sporadic outbursing
activity of high velocity neutron stars (NSs) in the extended
galactic halo with a distance scale of about $\sim 100-300$ kpc
(Shklovskii and Mitrofanov 1985, Li and Dermer 1992). High
velocity neutron stars are known to exist (Line and Lorimer
1994), and they have to escape the disk and to income into the
extended halo. Either starquakes or comets accretion are thought
to be energy resources of outbursts (for recent review see
Woosley 1996). And, each emitter has to provide a set of about
$\le 10^{4}$ bursts with the average energy $\sim 10^{42}$ erg
at each event.  

During a burst duration $\sim T_{d}$, the relativistic ejection
of a hot plasma occurs from a surface of NS into the
magnetosphere, which radiates a pulses of gamma-rays with
energies $\sim E_{p}$ and with duration from milliseconds up to
several tens of seconds.  After a {\it burst} stage, a NS could
go into a {\it relaxation} stage, which could be as long as a
post-glitch relaxation of radio pulsars. It is known to be about
several days (Cordes 1983). The energy release at this stage
could be much smaller than the total energy of the burst stage.
At the relaxation stage high energy photons could be radiated
with energies up to several GeV, and the surface of NS could be
heated up to $\sim$keV temperatures by discharges of sparks in
the magnetosphere. The post-burst relaxation stage might result
to afterglow emission as long as several days.  

\section{Testing the cosmological and galactic paradigms}

\vspace{2mm}

Different tests for both paradigms were proposed. Some of them
are decisive tests, which would provide the ensure conclusion
about the origin of GRBs.  Another tests could give more or less
preference to one or another model.  

\subsection{Direct astronomical identification of bursts emitters}

\vspace{2mm}

It would be distinctively positive for cosmological models, if
no-host problem will be resolved by direct measurements of
optical counterparts with cosmological red-shifts (see Metzger
{\it et al.} 1997). On the other hand, it would be distinctively
positive for galactic models, if proper motion of the optical
counterpart will be confirmed (Caraveo {\it et al.} 1997), or
some another direct identification with the galactic object will be
performed. Astronomical identification is the most conclusive
test to make a choice between two alternative models. 

\subsection{Bursts repetition}

\vspace{2mm}

The fact of bursts repetition, provided it would be found,
should be distinctively negative for cosmological models because
cosmological sources can not be recurrent, and it would
distinctively points out on the galactic halo emission. The
events like the set of four bursts of October 27-29, 1996
(Meegan {\it et al.} 1997b), have to be studied in more details
for the repetition test. 

\subsection{Spectral line-like features}

\vspace{2mm}

The finding of spectral line-like features would be
distinctively negative for cosmological models because fireball
emission can not have any narrow spectral lines. The presence of
lines would be positive (or neutral) for models which put bursts
in the extended halo. 

\subsection{Isotropy on the sky}

\vspace{2mm}

The perfect isotropy on the sky is very strong argument for the
cosmological model, but any significant deviation from the
isotropy would lead to rejection of this paradigm at all. The
galactic model predicts small dipole and/or quadruple effects
due to perturbations from the Milky Way and the Andromeda. The
detection of the extended halo around the Andromeda would be the
distinctively positive test for the galactic model.  

\subsection{Afterglow emission} 

\vspace{2mm}

Detection of bursts afterglow is the positive argument for the
cosmological model, provided the afterglow decrease would be
confirmed like $\sim t^{-1}$. For GRB 970508 some groups see
these kind of decay (Djorgovski {\it et al.} 1997), but some
another detects more rapid exponential decay (Kopylov {\it et
al.} 1997). On the other hand, the models of extended halo can
explain the afterglow, as the post-burst relaxation stage of
neutron star.  

\subsection{Time stretching of dim bursts}

\vspace{2mm}

All cosmological models predicts time-stretching of dim bursts
due to expansion of the Universe, and, therefore, detection of
time stretching would be a strong argument for these models.
However, one has to be sure that a found effect of
duration-brightness anti-correlation is actually resulted from
the cosmological time stretching.  

Internal evolution of sources might lead to this effect too.
The distinctive detection of cosmological time-stretching should
be based on several physically indepemdent {\it clocks} of
bursts, which would manifest the same time-dilation when
compared between bright and dim bursts.
On the other hand, some internal correlation between
luminosity and time of emission could easily explain time
stretching effects for sources in the galactic halo, and one
cannot consider this effect, as the conclusive test between two
models. 

\subsection{Red shifting of dim bursts}

\vspace{2mm}

The effect of hardness-brightness correlation of GRBs could be
interpreted as the evidence for cosmological red-shifting.
However, the groups of bright and dim bursts were found to have
quite similar trends of ($\alpha, E_{p})$ evolution, but the
evolution curve for dim group is shifted to larger $\alpha$ and
smaller $E_{p}$ values. It seems that the difference between dim
and bright bursts cannot be explained by a cosmological
transformation of standard candles, and some spectral evolution
of sources is required.  On the other hand, the effect could be
easily explained by the galactic model also, provided the
evolution of emitters would be assumed (Mitrofanov {\it et al.}
1997d). Therefore, one might conclude that the effect of
hardness-brightness correlation does not allow to make a
distinct choice between cosmological and galactic models, but
leads to the conclusion that emitters cannot be treated as {\it
standard candles}. 

\section{Potential developments in the Next Decade}

\vspace{2mm}

There are three main directions of potential development of
burst studies which could be predicted in the Next Decade.  

\subsection{New places in the space}

\vspace{2mm}

If classical bursts are cosmological, the largest red-shifts
of dimmest bursts $Z_{out}$ has to become resolved by
future measurements. These data will provide the unique way for
direct studies of stars evolution at the early Universe.  

The sample of the brightest bursts with accurate positions has
to be used to combine the {\it collective sky map} of X-ray,
optical and radio objects around the bursts locations.
Enhancement of any particular objects at the collective map
could point out the best candidate for bursts identification. On
the other hand, the absence of any counterparts will provide the
upper limit $Z_{in}$ for the distance scale of the most close
emitters.  

If classical bursts have the galactic origin, they are emitted
from the extended galactic halo. The similar extended halo
should be found around the Andromeda by more sensitive
instruments of the Next Decade. Direct comparison between bursts
from both halos will test bursts emission models. Large part of
high velocity neutron stars has to escape from the extended halo
around galaxies and give rise to the {\it Local group halo}
around the Local Group (Schaefer 1996). Very sensitive
instruments could detect bursts from neutron stars of
supergalactic halo provided they are still active having such a
large age.  

The future instruments are necessary with much higher
sensitivity for bursts detection down to $\sim
10^{-9}$~ergs/sm$^{-2}$ and with much better angular resolution
up to arc seconds. To achieve this reqirements, the X-ray all-
sky cameras would be the best facilities to perform the
continues observations, which could provide the on-board
triggering for each angular pixel of the imaging sky map.  

\subsection{New domains in the time}

\vspace{2mm}

Both for galactic and cosmological paradigms, the bursts energy
is thought to be released within the small volume of neutron
star or stellar black hole with the length of $R_{min}\sim
10^{6}$~cm. Therefore, the physical limit for bursts time scale
is $t_{min}\sim R_{min}/c \sim 30 \mu$s. Present instruments do
not see any bursts shorter than tens of milliseconds. On the
other hand, the distribution of bursts over parameter $T_{epw}$
shows that there could be a large number of short events with
$T_{d}<64$~ms.  

The future instruments are necessary with fast time resolution
up to microseconds. They probably should use the segmentation of
detectors and the logic of coincidence to detect events at
microseconds time scale.  

\subsection{New ranges of photons energy}

\vspace{2mm}

There are about 80 high energy photons which were detected by
EGRET from 5 bright BATSE GRBs at high energy range from tens of
MeV up to several GeV (Dingus 1995). The emission was seen from
$\sim$400~s before burst, as for GRB 910601, up to $\sim$ 5000~s after
it, as for GRB 940227. The key questions arises, how broadly is
high energy emission spreaded around a burst itself? Is emission
associated with the brightest bursts only, or with all of them?

Recent discoveries of the X-ray, optical and radio afterflow
emissions of GRBs gave very high priority to observations at the
low energy domain. The key question has to be responded, could
the low energy emission be seen before and during bursts, as
well as it is seen after them? Bursts with afterglow do not have
the same ratio between the fluence at gamma-rays and intensity
at another spectral ranges, so one needs to estimate the generic
relationship between fluence at different spectral ranges.  

The correlative program of future space/ground observations is
necessary with the fastest follow-up measurements at
different energy ranges.  The BeppoSAX mission has started this
program, and the Compton/Rossi duet is performing it now. The
future missions should be developed taking into account the
ability to participate in these correlative programs.  

 \section{Conclusion}

\vspace{2mm}

The studies of a particular burst could lead to understanding of
the phenomenon, only provided each event represents the all set.
I think that in the case of GRBs the situation could be quite
different: there is a broad variety of individual events which
have physically different, sometime even conflicting properties.

As well as quasars would not be selected within ordinary stars
using the optical plates only, new observations of gamma-ray
bursts are necessary, as suggested above, to fix the basic
properties of bursts, as the {\it necessary and sufficient}
conditions to identify the phenomenon. In parallel, a distinct groups of
bursts could probably be resolved with different, even
inconsistent basic properties. 

Taking into account well-known contradictions in the
observational properties of individual bursts, as presented
above, one could expect
that in the future studies the existence of two or more distinct
groups of bursts could probably be approved. The concepts of
{\it typical emitters} of bursts would be developed for these
groups, and theoretical models could be suggested to explain the
origin of separate classes of events.  

I think that it would be the main direction of studies of GRBs
for the Next Decade.  

\section{Acknowledgments}

\vspace{2mm}

The author would like to thank very much the Workshop Organizing
Committee for kind and warm hospitality.

\label{last}

\end{document}